\documentclass[10pt]{article}
\usepackage{graphicx}
\usepackage{amsmath}

\def\Title#1{\begin{center} {\Large #1 } \end{center}}
\def\Author#1{\begin{center}{ \sc #1} \end{center}}
\def\Address#1{\begin{center}{ \it #1} \end{center}}

\newcommand\pubblock{\rightline{\begin{tabular}{l} Proceedings of the Fifth Annual LHCP\\ \pubnumber\\
         \pubdate  \end{tabular}}}

\newenvironment{Abstract}{\begin{quotation} \begin{center} 
             \large ABSTRACT \end{center}\bigskip 
      \begin{center}\begin{large}}{\end{large}\end{center} \end{quotation}}

\newenvironment{Presented}{\begin{quotation} \begin{center} 
             PRESENTED AT\end{center}\bigskip 
      \begin{center}\begin{large}}{\end{large}\end{center} \end{quotation}}

\def\beq{\begin{equation}}
\def\eeq#1{\label{#1}\end{equation}}
\def\eeqn{\end{equation}}

\def\beqa{\begin{eqnarray}}
\def\eeqa#1{\label{#1}\end{eqnarray}}
\def\eeqan{\end{eqnarray}}

\let\bar=\overbar

\def\Dslash{\not{\hbox{\kern-4pt $D$}}}
\def\dslash{\not{\hbox{\kern-2pt $\del$}}}

\def\msb{{\bar{\ssstyle M \kern -1pt S}}}

\usepackage{color}

\newcommand\pubnumber{ CMS CR-2017-XXX }

\newcommand\pubdate{\today}

\def\affiliation{
On behalf of the CMS Collaboration, \\
Department of Physics and Astronomy \\
Ghent University, Ghent, 9000, Belgium}

\newcommand{\ttbar}{\ensuremath{\text{t}\bar{\text{t}}}}
\newcommand{\ttG}{\ensuremath{\ttbar\gamma}\,}
\newcommand{\ttH}{\ensuremath{\ttbar}H\,}
\newcommand{\ttZ}{\ensuremath{\ttbar}Z\,}
\newcommand{\ttW}{\ensuremath{\ttbar}W\,}

\newcommand{\ttX}{\ensuremath{\ttbar}X\,}
\newcommand{\fbinv}{\ensuremath{fb^{\rm -1}}\,}
\newcommand{\pt}{\ensuremath{\rm p_{T}}}

\begin{document}

\large
\begin{titlepage}
\pubblock

\vfill
\Title{ Studies of \ttX at CMS  }
\vfill

\Author{ Illia Khvastunov  }
\Address{\affiliation}
\vfill
\begin{Abstract}

This talk is dedicated to the most recent measurements of the processes that feature the production of a top quark pair either with an electroweak standard model boson, or another pair of top, bottom or light quarks. For the measurement of the production cross-section of top quark pair in association with a photon, 19.7 \fbinv of proton-proton collision data collected by CMS detector at $\sqrt{s}$ = 8 TeV is used, while for four top production, \ttbar + bb,  \ttbar + jj, \ttW and \ttZ the data collected at $\sqrt{s}$ = 13 TeV amounting to 35.9 \fbinv, is used. The measurement of \ttG is performed in the fiducial phase space corresponding to the semileptonic decay chain of the top quark pair, and the cross section is measured relative to the inclusive top quark pair production cross section. The fiducial cross section for this process is found to be  127 $\pm$ 27 (stat + syst) fb. The most recent search for the four top quark production, that explores the same-sign dilepton final state, set the upper observed (expected) limit to 4.6 ($2.9^{+1.4}_{-0.9}$) times predicted standard model cross section. The measured cross section of the top pair production with two b quarks is found to be $\sigma(\ttbar + bb)$ = 3.9 $\pm$ 0.6 (stat) $\pm$ 1.3 (syst) pb in the full phase space, while the production rate for the top pair production with two light quarks is measured to be $\sigma(\ttbar + jj)$ = 176 $\pm$ 5 (stat) $\pm$ 33 (syst) pb, and subsequently the ratio of the two is 0.022 $\pm$ 0.003 (stat) $\pm$ 0.006 (syst). The measurement of the \ttW and \ttZ processes combines three final states with two same-sign, three and four leptons. The \ttW and \ttZ production cross sections are measured to be $\sigma$(\ttZ) = $\rm 1.00^{+0.09}_{-0.08} (stat.)^{+0.12}_{-0.10} (sys.)$ pb and $\sigma$(\ttW) = $\rm 0.80^{+0.12}_{-0.11} (stat.)^{+0.13}_{-0.12} (sys.)$ pb with an expected (observed) significance of 4.6 (5.5) and 9.5 (9.9) standard deviations from the background-only hypothesis. The latter measurements are used to constrain the Wilson coefficients for four dimension-six operators describing interactions of new physics that would modify \ttW and \ttZ production.

\end{Abstract}
\vfill

\begin{Presented}
The Fifth Annual Conference\\
 on Large Hadron Collider Physics \\
Shanghai Jiao Tong University, Shanghai, China\\ 
May 15-20, 2017
\end{Presented}
\vfill
\end{titlepage}
\def\thefootnote{\fnsymbol{footnote}}
\setcounter{footnote}{0}
%

\normalsize 


\section{Introduction}

With the integrated luminosity collected at the Large Hadron Collider (LHC) in its first and second runs, rare processes producing top quark pairs in association with electroweak Standard Model (SM) bosons have become experimentally accessible. By measuring the cross section of top quark pairs produced in association with a photon or Z, the strength of the electromagnetic coupling of the top quark and photon or Z boson is probed. The top quark pair produced in association with a W or Z boson constitutes one of the heaviest set of SM particles that could be observed in the dataset accumulated so far. Also take note of the fact that all these processes are backgrounds for searches for new physics and for production of the Higgs boson with a pair of top quarks, the potential discovery of which became a hot topic in recent years. Any deviation from the measured cross section value from the SM prediction would be an indication of beyond the SM physics. Last but not least the cross section measurement of four top quark production provides a useful test of analytical higher order calculation of quantum chromodynamics (QCD). The top quark pair production in association with a photon was measured using the CMS detector\cite{CMSexperiment} using 19.7 \fbinv of proton-proton collision data collected at $\sqrt{s}$ = 8 TeV, while for all other processes described here the data volume collected at $\sqrt{s}$ = 13 TeV are used.

\section{The strategies of the analyses}
\subsection{Measurement of the \ttG cross-section}

For the \ttG cross section measurement\cite{tt_gamma} the final state was considered that consists of a high transverse momentum (\pt) lepton, missing transverse energy due to the presence of a neutrino, jets originating from both the b quarks (b jets) and from the decay of the W boson, and an energetic photon. The cross section of the process in this analysis is measured relatively to the \ttbar\,production cross section. In order to measure the ratio of these two cross sections, it is essential to measure how many \ttbar\,and \ttG events are observed in data. After the photon selection is applied, over half of the events in simulation come from background processes, and not from \ttG. The two largest backgrounds are \ttbar\,events which have a nonprompt photon coming from jets in the event and V+$\gamma$ events. There is no a single variable which can sufficiently discriminate both of these backgrounds for the signal. The V+$\gamma$  background can be differentiated from \ttG events by trying to reconstruct a top quark in the event, however \ttbar\,events are very similar to the signal in this respect. Alternatively, the nonprompt photon from the \ttbar\,background will tend to be less isolated than the photons from the signal, but the photon isolation variable will not be able to distinguish the V+$\gamma$ background from \ttG events. In order to be able to distinguish both \ttbar\,and V+$\gamma$ background events, both of these methods are exploited and the results are combined to measure the number of the \ttG events observed in data.
\subsection{Search of the four top quark production}

Two searches\cite{tttt_small,tttt_big} for four top quark production is based on the data collected in 2015 and 2016. Data collected in 2015 correspond to 2.6 \fbinv, while the data volume collected in 2016 are approximately 15 times greater and corresponds to 35.9 \fbinv. The first analysis exploits the final states with 1 or 2 light charged leptons and high hadron activity, while the second one is done in the context of a SUSY search in the same-sign dilepton final state. For the former a boosted decision tree (BDT) was developed which combines information on the event, including jet properties and distinguishes between four top and \ttbar\,production; the dominant uncertainty comes from the theoretical prediction of the \ttbar\,cross section, which include uncertainties on renormalisation and factorisation scales and parton distribution functions. The analysis in the same-sign dilepton final states exploits several categories classified according to lepton \pt, charge of the leptons, hadron activity, missing transverse momentum, jet and b jet multiplicity.

\subsection{Measurement of the \ttbar+bb and \ttbar+jj production cross-section and the ratio between them}

This analysis\cite{ttbb} is looking into the final state with an opposite-sign lepton pair and high jet multiplicity. Two of jets are expected to come from semileptonic top quark decay and in 85\% cases they are anticipated to be the leading and sub-leading b tagged jets. The b tagging discriminator of the 3rd and 4th jets is used to separate the \ttbar+bb from \ttbar+jj. The analysis is systematically constrained by the uncertainties in jet energy corrections and resolution, the choice of MC generator and theoretical uncertainties.

\subsection{Measurement of the \ttW and \ttZ cross-section}

The \ttW and \ttZ cross-section measurements\cite{ttV} are performed using events in which at least one of the W bosons originating from a top quark decays to a lepton and a neutrino, and the associated W boson decays to a lepton and a neutrino, or the Z boson decays to two charged leptons. Throughout this analysis only electrons and muons are considered as leptons, while $\tau$ leptons are only included through their leptonic decays. The \ttW process is measured in final states containing two leptons with equal charge. Requiring the same charge for the two leptons only the third of the signal produced in the dilepton final state will be selected. However this selection significantly improves the signal over background ratio, because prompt same-charge lepton pairs are only produced in SM processes with very small cross sections. The main backgrounds to this analysis are due to mis-reconstruction effects: misidentification of the so-called nonprompt leptons stemming from heavy flavour meson decays and electromagnetic charge mis-measurement for a prompt electron in an oppositely charged lepton pair. In order to distinguish these backgrounds from the signal, a multivariate analysis (MVA) has been developed. The preselection was applied to suppress the contribution from other backgrounds while not diminishing the signal significantly: 

\begin{itemize}
 \item the \pt\,of the two same-charge leptons should be greater than 25 GeV, in case of dielectron pair the highest-\pt(leading) and second highest-\pt(sub-leading) leptons have to pass 40 and 27 GeV, respectively, to ensure a high efficiency of event triggering;
 \item the final state is accompanied by a high multiplicity of jets and b jets, therefore at least 2 jets and 1 b jet is required in the selection;
 \item in order to suppress a Z boson to electron-positron pair events, the invariant mass of the two electrons is required to lie outside of a 15 GeV window around the Z pole mass;
 \item additionally to the previous, a requirement on transverse missing energy is added to completely suppress the contribution from Drell-Yan processes.
\end{itemize}

The MVA has been trained with ttW and \ttbar\,events as signal and background samples, respectively. A BDT classifier with a gradient boost was used and events were equally split for training and testing. Figure \ref{fig:figTTWbdt} shows the output of the BDT classifier in all background sources and the signal scaled to the integrated luminosity of the analysed data samples. The cut on BDT $>$ 0 value was used to suppress the background due to nonprompt leptons, while the BDT $<$ 0 region was used to constrain the uncertainty on the nonprompt lepton background estimation. For the final signal extraction the events were split into two categories: 0 $<$ BDT value $<$ 0.6 and BDT value $>$ 0.6. Furthermore to profit from the asymmetric yield in electron charge, the events are separated according to the charge of the leptons. The number of jets and b tagged jets are used as well to form exclusive event categories to maximize the signal significance.

\begin{table}[t]
\begin{center}
 \includegraphics[width=.4\textwidth]{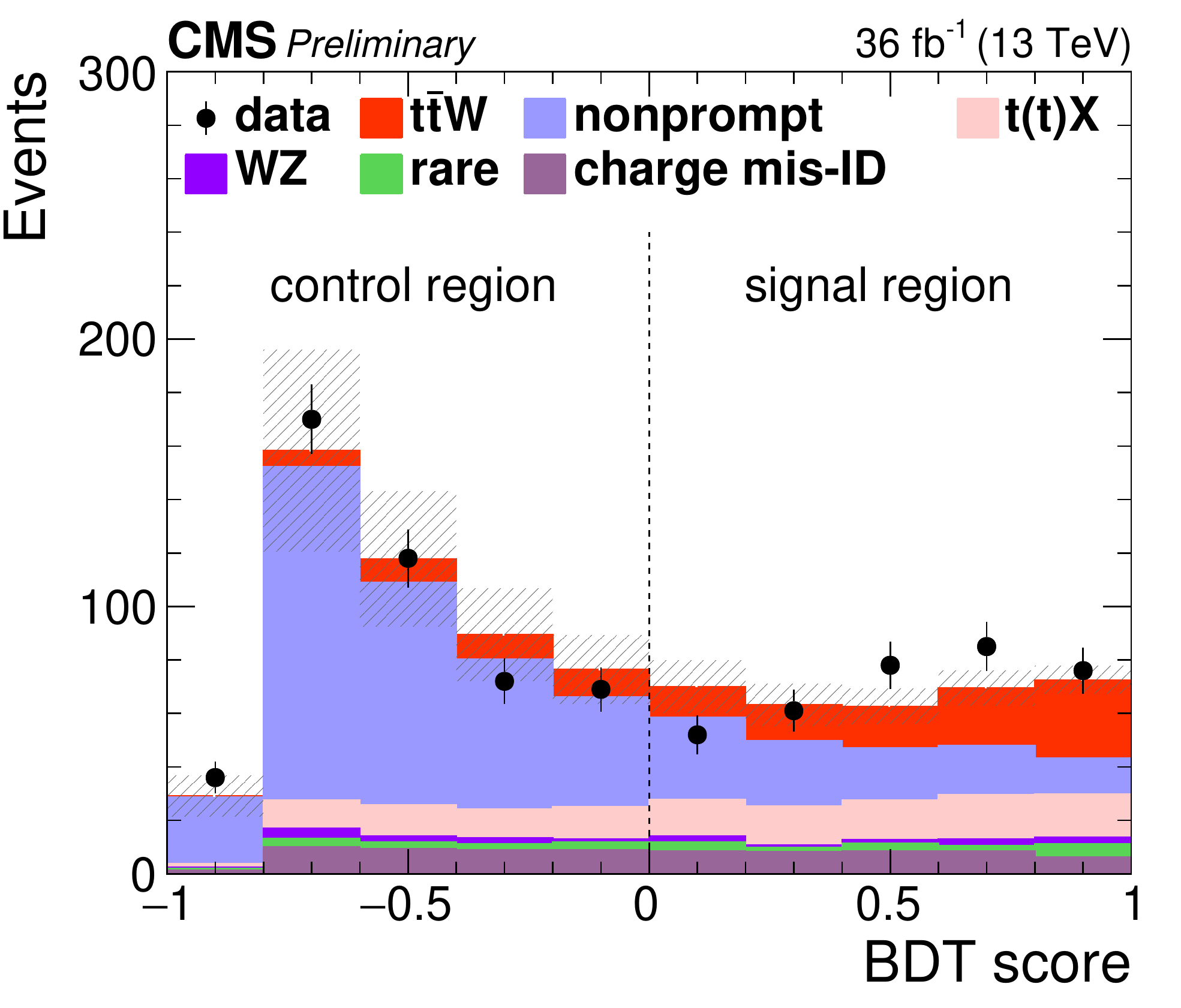}
\caption{ BDT value distribution for background and signal processes. The expected contribution from the different background processes are stacked as well as the expected contribution from the signal. The shaded band represents the uncertainty in the prediction of the background and the signal processes. Taken from \cite{ttV}.}
\label{fig:figTTWbdt}
\end{center}
\end{table}

The \ttZ proccess is measured in channels with three and four leptons, with a pair of same-flavour leptons of opposite charge, with an invariant mass close to the Z boson mass. In the three leptons channel the leading, sub-leading and lowest-\pt(trailing) transverse momenta has to pass 40, 20 and 10 GeV requirement, respectively. Additionally to that the cut on invariant mass of two same-flavour opposite-charge leptons and a cut on at least 2 jets are exploited. The same as in same-charge dilepton channel the number of jets and b tagged jets are used to form event categories to maximise the signal significance. In the four lepton channel, the leading lepton \pt is required to be greater than 40 GeV, while the transverse momenta of the remaining three leptons is required to be at least greater 10 GeV. The cut on the invariant mass of two leptons is exploited, as well as the cut on at least 2 jets. Events are divided into two exclusive categories according to number of b tagged jets.

\section{Results}

The number of events passing the photon selection containing top quark pairs can be measured by reconstructing the hadronic decaying top quark in the event. The M3 variable, defined as the invariant mass of the three jet combination that has the highest summed transverse momentum, is used for this purpose. The photon purity variable, defined as the fraction of reconstructed photons in the selection region which come from isolated photons as opposed to fake photons originating from jets, is used to discriminate between the types of real photons expected from signal and the hadronic produced photons from the \ttbar\,background. The fits for extracting the top quark purity and the photon purity are performed sequentially, and then the values are used in a likelihood function, from which a fit is performed to extract the number of events in the selection which originate from the \ttG signal process. The statistical uncertainty in the number of signal events dominates the determination of the cross section for \ttG. It includes the uncertainties on the measurement of the photon purity, top purity after photon selection and the statistical uncertainty from the number of events in data. The ratio of the fiducial cross section of \ttG to \ttbar\,production, R, can be found in Table \ref{tab:tableTTG} for both electron+jets and muon+jets final state as well as fiducial cross section and the cross section times branching ratio. The measured cross section is in agreement with the theoretical prediction.

\begin{table}[t]
\begin{center}
 \includegraphics[width=.6\textwidth]{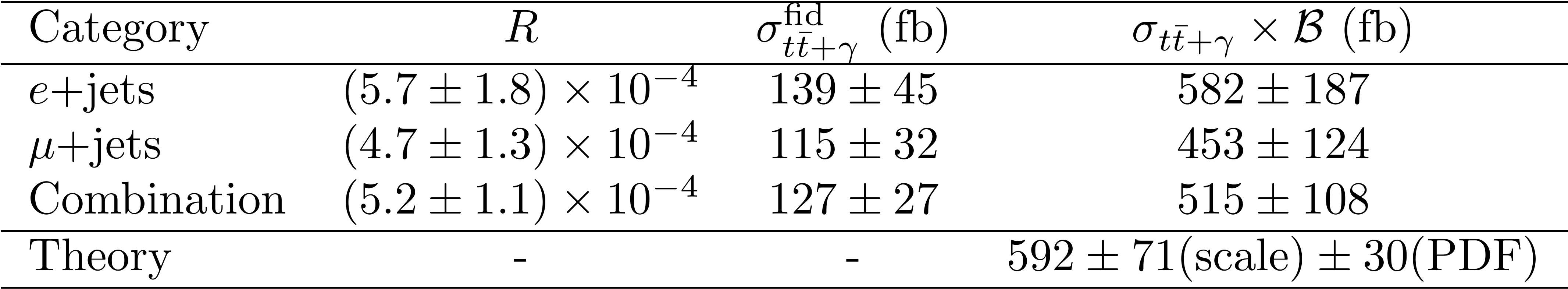}
\caption{ Summary of the measured cross section for \ttG process in electron + jets, muon + jets final states and the combination of the two: the ratio of the fiducial cross section of \ttG to \ttbar\,production, R (first column), fiducial cross sections for \ttG process and the measured cross section times branching ratio. Taken from \cite{tt_gamma}.}
\label{tab:tableTTG}
\end{center}
\end{table}

The cross section for the production of \ttbar+bb and \ttbar+jj events and their ratio are found to be $\sigma(\ttbar + bb)$ = 3.9 $\pm$ 0.6 (stat) $\pm$ 1.3 (syst) pb, $\sigma(\ttbar + jj)$ = 176 $\pm$ 5 (stat) $\pm$ 33 (syst) pb and 0.022 $\pm$ 0.003 (stat) $\pm$ 0.006 (syst). The measurements are compared with the standard model expectations obtained from a POWHEG simulation at the next-to-leading-order interfaced with PYTHIA.

Both four top searches improved previous round of searches and the observed (expected) limit to 4.6 ($2.9^{+1.4}_{-0.9}$) SM cross sections. 

\begin{figure}[h!]
\centering

  \includegraphics[width=.65\textwidth]{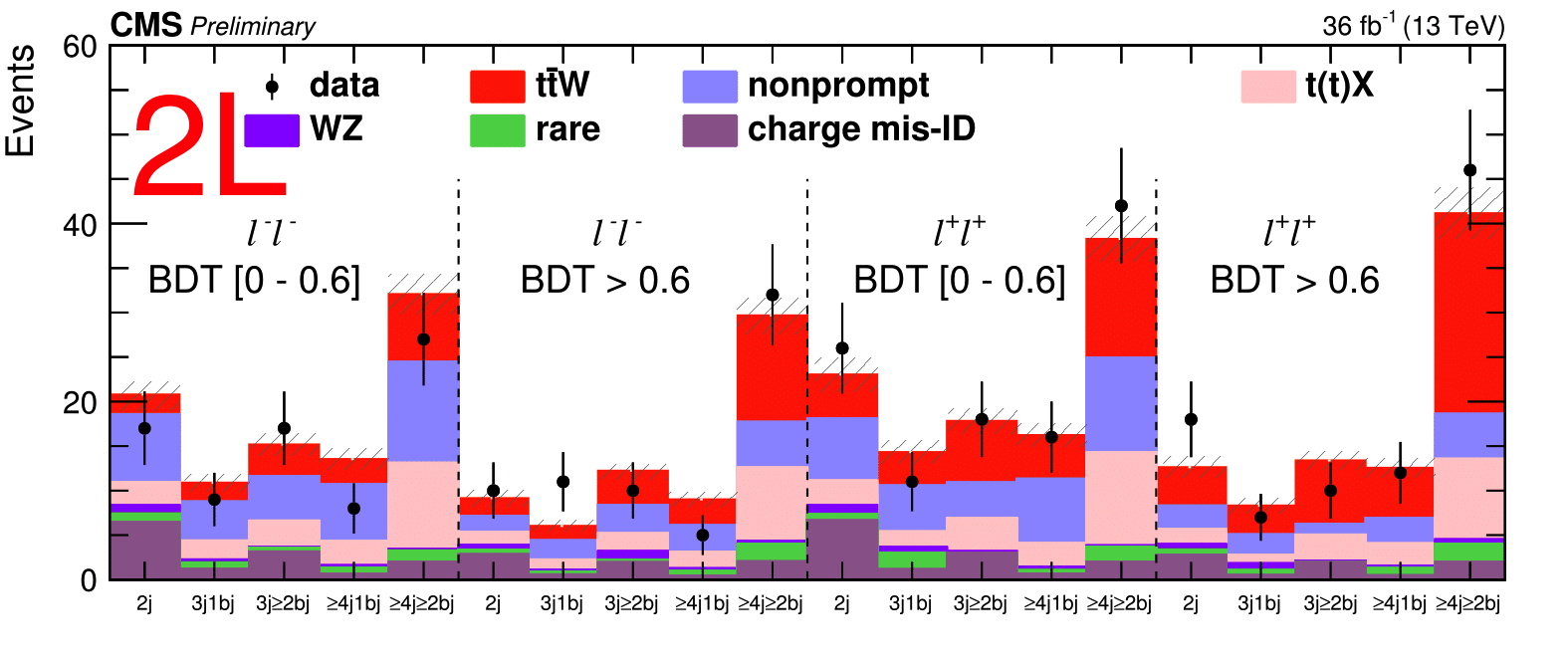}
  \includegraphics[width=.49\columnwidth]{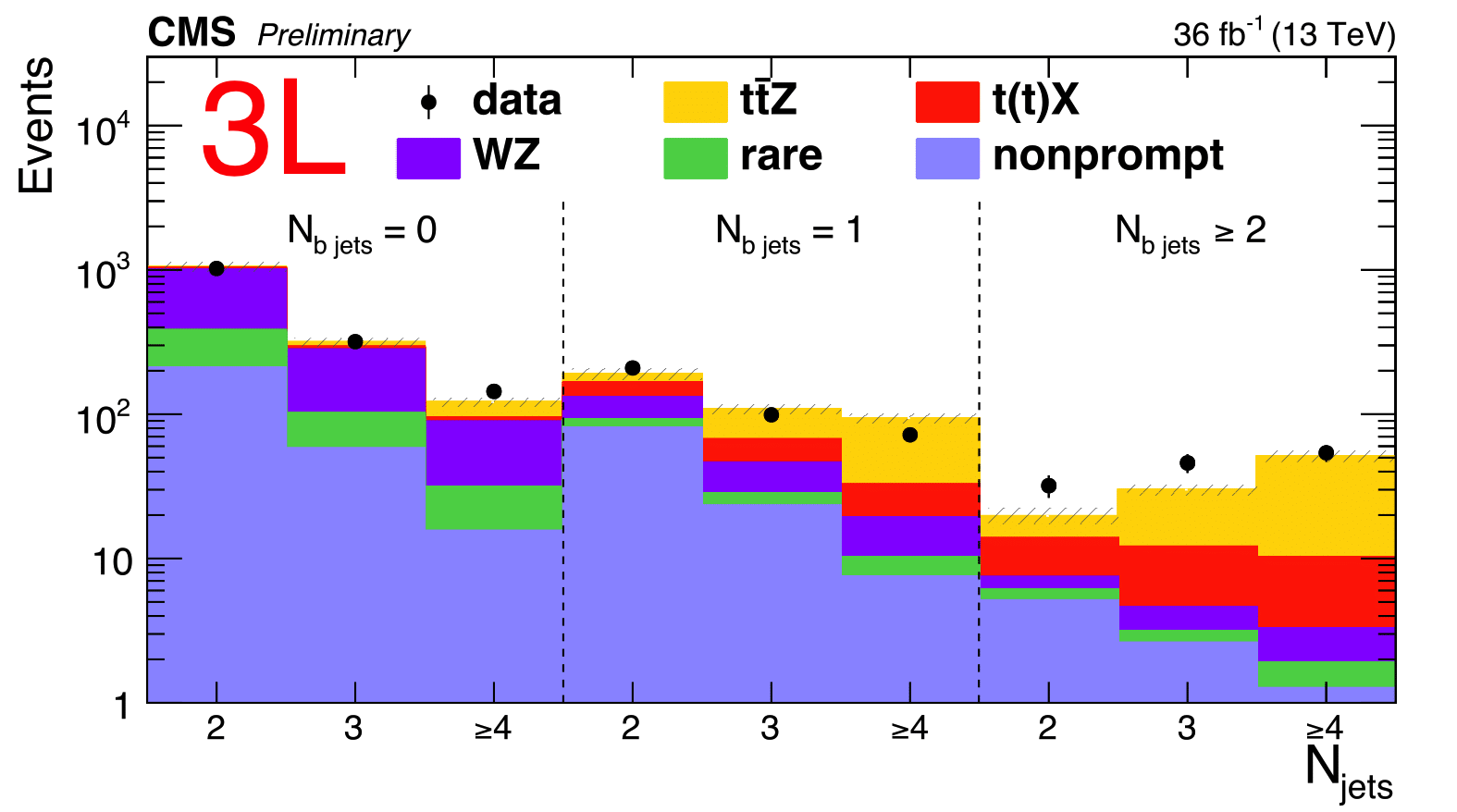}
  \includegraphics[width=.29\columnwidth]{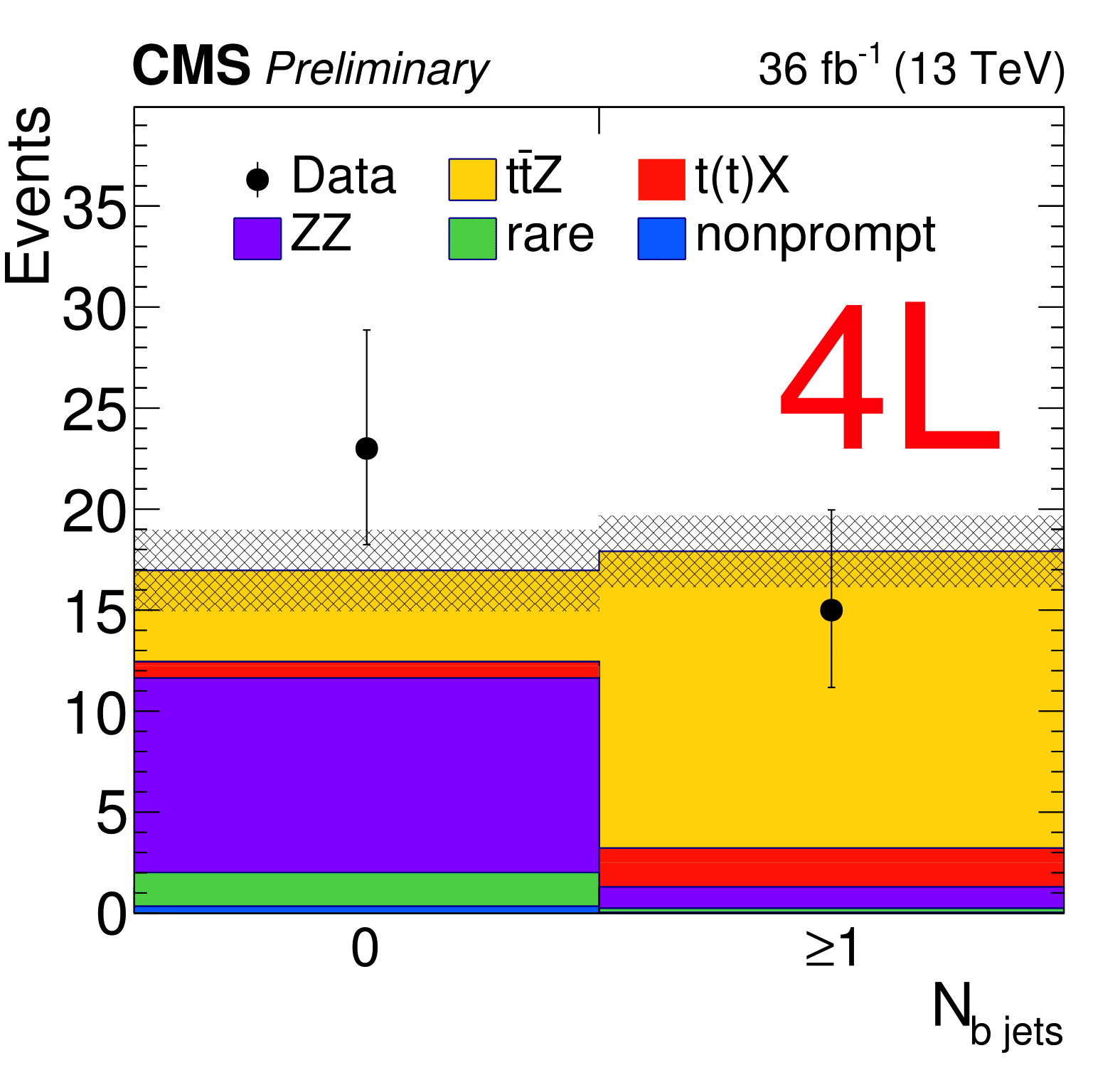}

\caption{ Predicted signal and background yields, compared to observed data in the same-sign dilepton (upper), trilepton (bottom left) and four-lepton (bottom right) analyses. The hatched band shows the total uncertainty associated with the signal and background predictions. Taken from \cite{ttV}.}
\label{fig:figureTTV}
\end{figure}

A binned maximum likelihood fit is performed over all the signal regions to measure the production cross sections of \ttW and \ttZ. Systematic uncertainties are treated as nuisance parameters for the fit. The greatest effect both on the \ttW and \ttZ cross section measurement is arising from lepton reconstruction, b tagging mis-modeling and trigger efficiency measurement. The uncertainty on nonprompt background gives a significant contribution to the systematic uncertainty of \ttW cross section measurement. Overall with the data collected in 2016 the systematic uncertainty for \ttW and \ttZ cross section measurement becomes dominant. The estimated yields and the observed number of events are shown for dilepton, trilepton and four-lepton final states on Figure \ref{fig:figureTTV}. 

\begin{figure}[h!]
\centering

  \includegraphics[width=.5\textwidth]{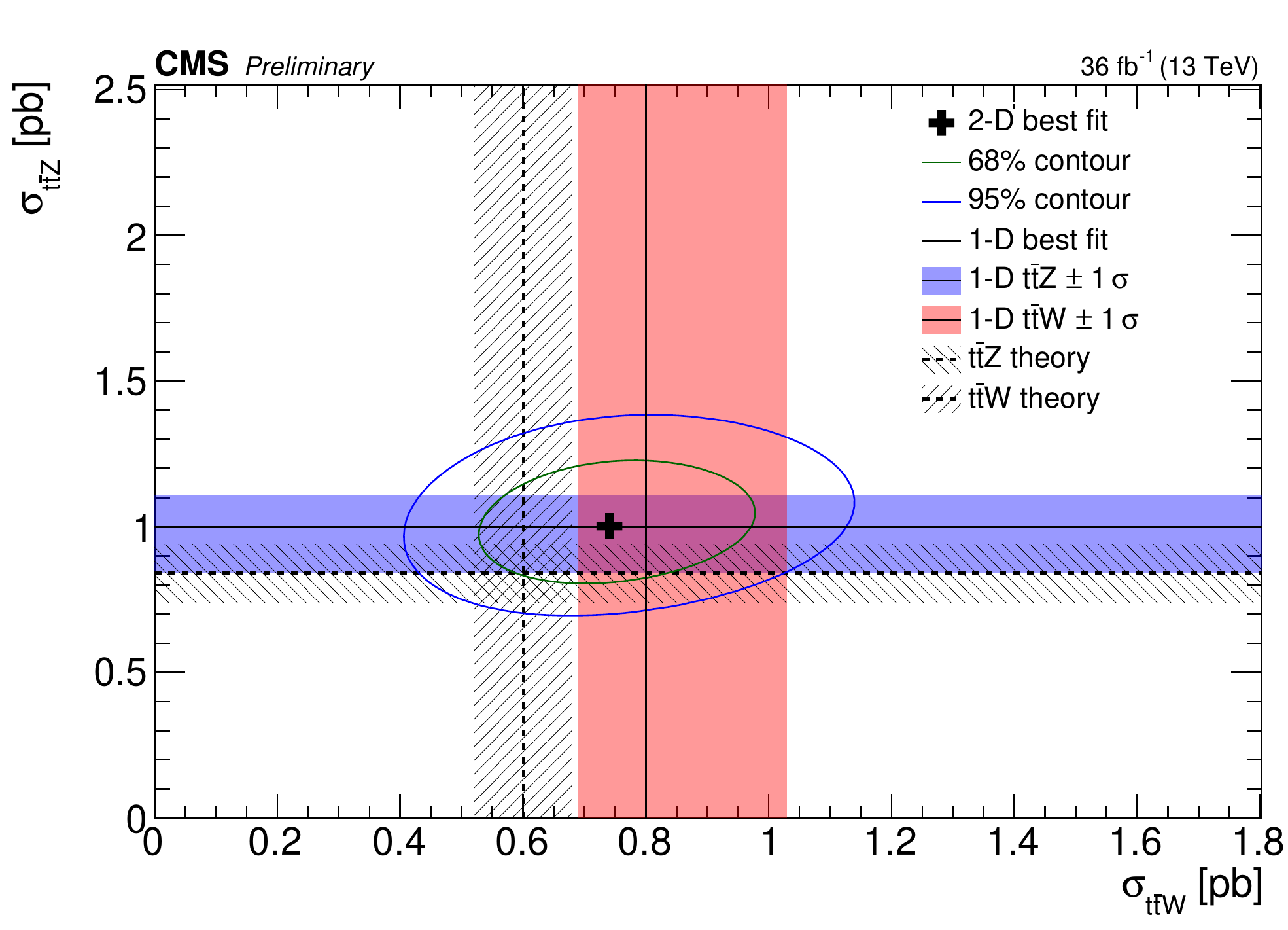}

\caption{ The result of the two-dimensional best fit for \ttW and \ttZ cross sections (cross symbol) is shown along with its 68 and 95\% confidence level contours. The result of this fit is superimposed with the separate \ttW and \ttZ cross section measurements, and the corresponding 1$\sigma$ bands, obtained from the dilepton, and the three lepton/four-lepton channels, respectively. The figure also shows the predictions from theory and the corresponding uncertainties. Taken from \cite{ttV}.}
\label{fig:figureTTVFit2D}
\end{figure}

The results of the individual fits are summarised in Figure \ref{fig:figureTTVFit2D} as well as a simultaneous fit of the cross sections of the two processes using all dilepton, tri-lepton, and four-lepton channels at the same time. The cross section extracted from this two-dimensional fit for ttZ is identical to those obtained from the two one-dimensional fit, while for ttW the two-dimensional fit is shifted down by approximately 9\% towards the theoretical prediction. The two-dimensional fit is in agreement with the theoretical prediction within 1$\sigma$ band. 

Under effective field theory, cross section measurements can be used to search in a model-independent way for new physics at energy scales which are not yet experimentally accessible. Using this approach, the SM Lagrangian is extended with higher-order operators which correspond to different combinations of SM fields.  In the analysis the couplings to the first two generators, as well operators which caused significant cross section scaling for \ttbar, inclusive Higgs, WW and WZ are not considered. The chosen ones are the operators that affects the \ttH, \ttW and \ttZ cross sections. The expected confidence level intervals for the selected Wilson coefficients, which parameterise the strength of the new physics interaction, are summarised in the Table \ref{tab:tableTTVEFT}.

\begin{table}[t]
\begin{center}
  \includegraphics[width=.7\columnwidth]{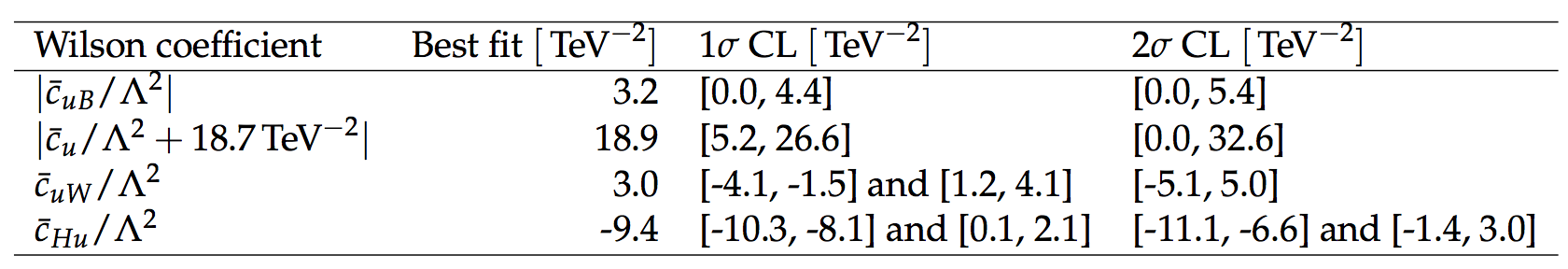}
\caption{ Observed best-fit values determined from \ttW and \ttZ measurement, along with corresponding 1$\sigma$ and 2$\sigma$ confidence level intervals for selected Wilson coefficients. Taken from \cite{ttV}.}
\label{tab:tableTTVEFT}
\end{center}
\end{table}


\end{document}